\documentclass[12pt,notitlepage,a4paper]{article}
\usepackage{color,graphicx}
\usepackage{cite}
\usepackage{amssymb}
\usepackage{delarray,amsmath}
\usepackage{mathtools}
\usepackage[table]{xcolor}
\usepackage[top=30mm, bottom=30mm, right=25mm, left=25mm, headheight=15.72pt]{geometry}
\usepackage{hyperref}

\usepackage[utf8]{inputenc}
\usepackage[T1]{fontenc}

\newcommand{\be}{\begin{equation}} \newcommand{\ee}{\end{equation}}
\newcommand{\bea}{\begin{eqnarray}} \newcommand{\eea}{\end{eqnarray}}
\newcommand{\beann}{\begin{eqnarray*}}  \newcommand{\eeann}{\end{eqnarray*}}
\newcommand{\bfig}{\begin{figure}} \newcommand{\efig}{\end{figure}}
\newcommand{\ba}{\begin{array}} \newcommand{\ea}{\end{array}}
\newcommand{\bcen}{\begin{center}} \newcommand{\ecen}{\end{center}}
\newcommand{\btab}{\begin{tabular}} \newcommand{\etab}{\end{tabular}}

     \def\sign{\operatorname{sign}}
   
\newcommand{\vev}[1]{\left\langle{#1}\right\rangle}

\newtheorem{Proposition}{Proposition}[section]

\newtheorem{Theorem}{Theorem}[section]
\newtheorem{Lemma}{Lemma}[section]
\newtheorem{Corrolary}{Corrolary}[section]

\newcommand{\bp}{\begin{Proposition}}   \newcommand{\ep}{\end{Proposition}}
\newcommand{\bt}{\begin{Theorem}}   \newcommand{\et}{\end{Theorem}}
\newcommand{\bl}{\begin{Lemma}}     \newcommand{\el}{\end{Lemma}}
\newcommand{\bc}{\begin{Corrolary}} \newcommand{\ec}{\end{Corrolary}}

\def\m{\mu}

\usepackage{amsthm}
\usepackage{bm}

\numberwithin{equation}{section}

\title{
\bf{Gapped dilatons in scale invariant superfluids}}
\date{}
\author{Riccardo Argurio$^{a,}$\footnote{rargurio@ulb.ac.be}, 
Carlos Hoyos$^{b,}$\footnote{hoyoscarlos@uniovi.es}, 
Daniele Musso$^{c,}$\footnote{daniele.musso@usc.es}~ and 
Daniel Naegels$^{a,}$\footnote{daniel.naegels@ulb.ac.be}}

\begin{document}
\maketitle
\begin{center}\it{
$^{a}$Physique Th\'eorique et Math\'ematique and International Solvay Institutes, \\ Universit\'e Libre de Bruxelles, C.P. 231, B-1050 Brussels, Belgium\\
\vspace{15pt}
$^{b}$Department of Physics and\\ 
Instituto de Ciencias y Tecnolog\'ias Espaciales de Asturias (ICTEA)\\
 Universidad de Oviedo,  c/ Federico Garc\'{\i}a Lorca 18, E-33007 Oviedo, Spain\\
\vspace{15pt}
$^{c}$Departamento de F\'\i sica de Part\'\i culas and\\
Instituto Galego de F\'\i sica de Altas Enerx\'\i as (IGFAE)\\
Universidade de Santiago de Compostela, E-15782 Santiago de Compostela, Spain\\
Inovalabs Digital S.L. (TECHEYE), E-36202 Vigo, Spain}
\end{center}
\vspace{25pt}

\begin{abstract}
\noindent
We study a paradigmatic model in field theory 
where a global $U(1)$ and scale symmetries are jointly and 
spontaneously broken.
At zero density the model has a non-compact flat direction, which at finite density needs to be slightly lifted.
The resulting low-energy spectrum is composed by a standard 
gapless $U(1)$ Nambu-Goldstone mode and a 
light dilaton whose gap is determined by the chemical 
potential and corrected by the couplings. 
Even though $U(1)$ and scale symmetries commute, there is a 
mixing between the $U(1)$ Nambu-Goldstone and the dilaton that is crucial to 
recover the expected dynamics of a conformal fluid and leads
to a phonon propagating at the speed of sound.
The results rely solely on an accurate study of the 
Ward-Takahashi identities and are checked against standard 
fluctuation computations. We extend our results to a boosted superfluid, and comment the 
relevance of our findings to condensed matter applications.
\end{abstract}
\newpage
\tableofcontents

\section{Introduction}

Scale invariance plays a special role in many-body and high-energy physics.
It underlies the emergence of universality in many instances, such as critical phenomena, 
Landau-Fermi liquids or cold atoms at unitarity, to name a few. 
Scale transformations are a symmetry either at very low or very high 
energies compared to the intrinsic scales. In most cases they represent 
only an approximate symmetry valid in a restricted regime, requiring 
typically a certain degree of fine-tuning in the interactions, the 
thermodynamic variables, the external parameters, or the support of 
additional symmetries.
When a scale invariant system is considered at non-zero particle number or
at finite charge density, scale symmetry is spontaneously broken; such 
breaking is directly relevant to characterize the dynamics of the system 
mentioned above but it can also be useful to extract properties of large 
charge operators of a CFT via the state-operator correspondence 
\cite{Hellerman:2015nra,Alvarez-Gaume:2016vff,Monin:2016jmo,Orlando:2019skh}.

Whenever an internal symmetry is spontaneously broken in a relativistic 
system, one expects to encounter gapless excitations in the form of 
Nambu-Goldstone (NG) modes \cite{Nambu:1960tm,Goldstone:1961eq,PhysRev.127.965}, 
one for each broken symmetry.
If instead the breaking involves spacetime symmetries, the counting of modes 
becomes more complicated \cite{Ivanov:1975zq,Low_2002,Watanabe_2013,Brauner_2014}, 
yet the presence of a Nambu-Goldstone mode associated to scale 
invariance, commonly known as \emph{dilaton}, is still a possibility. 
Similarly, the counting of NG modes deviates from the standard
Goldstone theorem expectation when the system is not Lorentz invariant
\cite{NIELSEN1976445,Watanabe:2011ec,Watanabe_2012,Kapustin:2012cr,Watanabe_2013}.%
\footnote{For a recent review on NG counting rules we refer to
\cite{Watanabe_2020}.} 

Even in relativistic systems, the presence of a non-zero charge density 
breaks the boosts spontaneously, thus the NG modes may 
show some features similar to those emerging in non-relativistic systems. 
In the case of several internal symmetries, there can be additional 
{\em gapped} modes besides the gapless NG modes
\cite{Miransky:2001tw,Schafer:2001bq,Nicolis_2013,Watanabe:2013uya,Nicolis:2013sga,Endlich_2015}.
More specifically, in the presence of a chemical potential $\mu$ for a 
conserved charge $Q$, gapped modes emerge when the effective Hamiltonian\footnote{We are proceeding in analogy to \cite{Nicolis_2013,Watanabe:2013uya,Endlich_2015,Arav_2017}.}  
$\widetilde{H}=H-\mu Q$ does not commute with the broken generators. The gap is fixed by group theory 
considerations and is proportional to the chemical potential.
In general, there can also be additional modes whose gap, although 
proportional to $\mu$, is not protected by symmetry \cite{Kapustin:2012cr,Nicolis:2013sga}. 
Such analysis was generalized in \cite{Watanabe:2013uya} to cases where $\mu Q$ 
in the effective Hamiltonian is replaced by some other deformation involving a 
symmetry generator, for instance the magnetic field times the spin in a 
ferromagnet $\widetilde{H}=H-g B_z S_z$. 

The breaking of scale invariance presents some similarities with the story 
above due to the fact that the generator of dilatations $D$ does not commute with the 
Hamiltonian $[D,H]=iH$.%
\footnote{Nevertheless, it is a conserved charge because $\partial_t D=H$, 
so its total time derivative in the Heisenberg picture vanishes.} 
Accordingly, the commutator with the effective Hamiltonian is
\be\label{eq:comHD}
[\widetilde{H},D]=-iH.
\ee
For simplicity let us assume that $Q$ is the generator of an Abelian 
$U(1)$ symmetry. If this symmetry is spontaneously broken, the ground
state is not an eigenstate of $Q$. However, it must be by definition an 
eigenstate of $\widetilde{H}$, so time-translations generated by $H$ are 
spontaneously broken too. In fact time translations and the $U(1)$ symmetry 
are broken to a diagonal subgroup and there is just a single NG
mode associated to both generators.

Equation \eqref{eq:comHD} implies that there is a mixing between the $U(1)$ 
NG and the dilaton, then --even though $Q$ commutes with 
$\widetilde{H}$-- the state produced by the corresponding charge density $J^0$
applied to the vacuum at some initial time is not an eigenstate of time evolution.
Borrowing an analogy from high-energy physics, the ``flavor'' eigenstates 
defined by the symmetry generators are not aligned with the ``mass'' eigenstates.
If the dilaton were not dynamical, or if it were integrated out, the mixing implied by \eqref{eq:comHD} would be manifested in the form of an inverse Higgs constraint.

Although interesting, one might wonder whether it is sensible to discuss the 
physics of a dilaton in the first place, since the energy density is in general 
non-zero at non-zero charge density. In that case, a scale transformation 
would change the vacuum energy density (as determined by the temporal component
of the energy-momentum tensor $T^{\mu\nu}$) by an amount proportional to itself
\be
\delta \vev{T^{00}} \sim \vev{ -i[D,T^{00}] }=(d+1)\vev{T^{00}},
\ee
where $d+1$ is the number of spacetime dimensions.
Both here and henceforth, we assume a relativistic theory, thus there cannot be 
a NG mode associated to the spontaneous breaking of scale invariance 
unless $\vev{T^{00}}=0$.%
\footnote{Note also that the combination of Lorentz invariance (which fixes 
the expectation value of the energy-momentum tensor to $\vev{T^{\mu\nu}}=\Lambda \eta^{\mu\nu}$) 
and the Ward-Takahashi identity for scale invariance, $\vev{T^\mu_{\ \mu}}=0$, fixes $\vev{T^{00}}=0$.}
This is quite restrictive. Since a gapless mode requires a degeneracy of ground states, 
the theory needs to have a moduli space of vacua in addition to scale invariance:
these are {\em flat directions} in the potential, supposing we refer to a field theory with a Lagrangian.%
\footnote{A related discussion about fine-tuning the cosmological constant to zero in order to have a flat dilatonic direction is contained in \cite{Amit:1984ri,Rabinovici:1987tf,Komargodski_2011,Chai:2020zgq}.}

Maybe contrary to expectations, the situation at finite density is similar despite the fact that the energy density is non-vanishing. 
If the ground state is homogeneous and isotropic, the expectation value of the components of the energy-momentum tensor 
correspond to constant energy density and pressure
\be
\vev{T^{00}}=\varepsilon,\qquad \vev{T^{ij}}=p\, \delta^{ij}\ .
\ee
Scale invariance implies that the expectation value of the trace of 
the energy-momentum tensor will vanish $\vev{T^\mu_{\ \m}}=0$, 
which fixes the equation of state $\varepsilon=dp$, where $d$ is the number of spatial dimensions.  In addition, 
we have the usual relation between thermodynamic potentials at zero 
temperature, $\varepsilon+p=\mu \rho$, where $\rho=\vev{J^0}$ is 
the $U(1)$ charge density. Combining the two, the energy density of
the scale invariant theory is $\varepsilon=d/(d+1)\mu \rho$. At 
finite density the relevant quantity is not the energy density, but
the free energy (density) given by the effective Hamiltonian 
$T^{00}-\mu J^0$. A scale transformation changes the expectation of 
the effective energy density as follows:
\be
\delta \vev{T^{00}-\mu J^0}\sim \vev{ -i[D,T^{00}] }-\mu \vev{ -i[D,J^{0}] }=(d+1)\varepsilon-d \mu \rho=0\ .
\ee
Then, under quite general assumptions, scale transformations do not 
shift the free energy of a finite density state in a scale 
invariant theory and it is legitimate to discuss the physics of a 
dilaton mode, at least at zero temperature.

The observation above does not imply directly the existence of a gapless (or gapped) mode. 
In the absence of a general argument that would allow us to fix the properties of a 
dilaton mode, we study a concrete model of spontaneous breaking of scale invariance at 
non-zero density. We restrict the analysis to a relativistic theory in $3+1$ dimensions, 
and keep the analysis classical.
Such simple model is informative because it can be interpreted as an effective action \`a la Ginzburg-Landau for the order parameter. 

The principal highlights of the 
present study are two. On one side, the characterization of the dilaton dispersion relation 
and particularly its gap. This concerns mainly the effects of the chemical potential and 
its role in defining the effective low-energy spectrum.
On the other side, we propose and check a method based uniquely on the study of Ward-Takahashi 
identities, that in our setup just correspond to classical conservation equations.

The paper is structured as follows. Section \ref{modello} introduces the model at zero density,
where we emphasize the need for flat directions in the potential. This condition is relaxed in 
Section \ref{muuu}, where we study the model at non-zero density. In Section \ref{vvv} 
the analysis is extended to allow for non-zero superfluid velocity. 
Each section has a subsection dedicated to the analysis of the 
Ward-Takahashi identities, together with a check of the latter method against standard Lagrangian 
computations for the fluctuations. We conclude the paper in Section \ref{disco} with further 
comments on the results, their interpretation, their applications and possible extensions.

\section{The Model}
\label{modello}

Consider the standard Goldstone model for a global $U(1)$ symmetry in 4 spacetime dimensions
\begin{align}\label{toytoy}
S=\int d^4x\ \left[ \partial_\mu \psi \partial^\mu \psi^* - \lambda (|\psi|^2-v^2)^2\right] \ ,
\end{align}
where $\psi$ is a scalar complex field charged under the $U(1)$ symmetry, which acts as $\psi \to e^{i\alpha}\psi$, 
while $\lambda$ and $v$ represent --respectively-- a dimensionless and a dimensionful coupling.
Given the presence of a dimensionful coupling, the model \eqref{toytoy} does not enjoy scale invariance. 
We can nonetheless make it scale invariant if we replace $v$ with a dynamical real scalar field $\xi$ acting 
as a compensator:
\begin{align}\label{tuned}
S=\int d^4x\ \left[ \partial_\mu \psi \partial^\mu \psi^* +\frac12\partial_\mu\xi\partial^\mu\xi- \lambda (|\psi|^2-\xi^2)^2\right] \ .
\end{align}

The equations of motion are given by
\begin{align}
\partial^2 \psi+2\lambda(|\psi|^2-\xi^2)\psi=0\ ,\qquad
\partial^2 \xi-4\lambda(|\psi|^2-\xi^2)\xi=0 \ ,
\end{align}
and the generic stationary solution is 
\begin{equation}
\xi = v\ , \qquad \qquad |\psi|^2=v^2\ .
\label{vacToyToy}
\end{equation}
The space of solutions \eqref{vacToyToy} has two moduli, $\xi$ itself and the phase of $\psi$. 
Consider the fluctuations around \eqref{vacToyToy}, parameterized as follows
\begin{equation}\label{psiparam}
\begin{split}
\psi&=e^{i\frac{\vartheta}{\sqrt2v}} \left(v\, e^{\frac{\tau}{\sqrt3v}}+\frac{\rho}{\sqrt6}\right)\ \simeq\  v+\frac{\tau}{\sqrt 3}+ \frac{\rho}{\sqrt6}+i\frac{\vartheta}{\sqrt2}\ , \\
\xi&=v\, e^{\frac{\tau}{\sqrt3v}}-2\frac{\rho}{\sqrt6}\ \simeq\ v+\frac{\tau}{\sqrt3}-2\frac{\rho}{\sqrt6}\ ,
\end{split}
\end{equation}
where $\tau$, $ \rho$ and $\theta$ are real.
The quadratic action for the fluctuations is given by
\begin{align}\label{qua_lor}
S_\mathrm{quad}=\int d^4x\ \left[\frac12\partial_\mu\tau\partial^\mu\tau+ \frac12\partial_\mu  \rho \partial^\mu  \rho 
+\frac12 \partial_\mu \vartheta \partial^\mu \vartheta -6 \lambda v^2 \rho^2\right] \ .
\end{align}
We thus see that $\rho$ gets a mass $12\lambda v^2$ while $\tau$ and $\vartheta$ are massless. We identify the latter two with the Goldstone bosons for broken scale invariance, the dilaton, and for broken $U(1)$ symmetry, the $U(1)$ NG.
The dispersion relations are trivially relativistic, since Lorentz symmetry is preserved.

In order to study the low-energy modes about \eqref{vacToyToy}, one can alternatively rely entirely on symmetry considerations and, specifically,
on the Ward-Takahashi identities. As we will show in the next subsection, such symmetry-aware approach permits to obtain the equations 
of motion for the low-energy modes in a direct way, which is usually more transparent than the standard Lagrangian 
study of the fluctuations.

\subsection{Ward-Takahashi identities and low-energy modes}

Model \eqref{tuned} features a conserved $U(1)$ current given by
\be
J_\mu=i(\partial_\mu\psi^*\psi-\psi^*\partial_\mu\psi)\ , \qquad \partial^\mu J_\mu=0\ ,
\ee
while the improved energy-momentum tensor is
\be
\label{tmunu}
T_{\mu\nu}=2\partial_{(\mu}\psi^*\partial_{\nu)}\psi+\partial_\mu \xi\partial_\nu\xi-\eta_{\mu\nu}{\cal L}+\frac{1}{3}(\eta_{\mu\nu}\partial^2-\partial_\mu\partial_\nu)\left(\frac{1}{2}\xi^2+|\psi|^2\right)\ .
\ee
This expression satisfies on-shell the following Ward-Takahashi identities
\footnote{The trace Ward-Takahashi identity requires the improvement introduced in \eqref{tmunu}.}
\be
T_{[\mu\nu]}=0\ ,\qquad \partial^\mu T_{\mu\nu}=0\ , \qquad T^\mu_{\ \mu}=0 \ .
\ee
We expand around the vacuum \eqref{vacToyToy} by considering the fluctuation parametrization \eqref{psiparam}.
Up to linear order in the fields, the $U(1)$ current is given by
\be
J_\mu\simeq \sqrt2v\partial_\mu \vartheta\ ,
\ee
so that its conservation equation gives the equation of motion for the $U(1)$ NG mode
\be
0=\partial^\mu J_\mu\simeq \sqrt2v \partial^2\vartheta \ .
\label{EOMPhasonToyToy}
\ee
The energy-momentum tensor expanded to linear order is
\be
T_{\mu\nu}\simeq \frac{v}{\sqrt3}(\eta_{\mu\nu}\partial^2-\partial_\mu\partial_\nu)\tau\ ,
\ee
and the trace Ward-Takahashi identity yields the equation of motion for the dilaton
\be
0=T^\mu_{\ \mu}\simeq \sqrt3v\partial^2\tau \ .
\label{EOMDilatonToyToy}
\ee
From \eqref{EOMPhasonToyToy} and \eqref{EOMDilatonToyToy} we can observe that we recover the two massless modes of \eqref{qua_lor}. 
The Ward-Takahashi computation, however, descends directly from symmetry arguments, being therefore more convenient (and easier) 
to apply, especially when dealing with models more complicated than \eqref{tuned}. In particular, this approach allows to identify immediately and without ambiguities the nature of each Goldstone boson, simply by associating every (gapless) mode to the Ward-Takahashi identity that yields its equation of motion.

It is important to stress that the model \eqref{tuned} is fine-tuned. 
Indeed, (classical) scale invariance dictates that the potential should contain only quartic terms in the scalars,
but the fact that the potential is a perfect square constitutes a fine-tuning, specifically considered to the purpose of having a flat direction. 
The latter is of course a necessary condition for the presence of a low-energy dilaton mode.

The simple argument is as follows. In such a relativistic set-up, scale invariance implies the absence of any reference scale in the (effective) Lagrangian. If scale invariance is to be broken spontaneously by a vacuum expectation value (VEV), then the latter must be arbitrary. Hence this VEV parameterizes a non-compact flat direction. Moreover the absence of any reference scale means that the flat direction must also correspond to a vanishing vacuum energy. The particle which corresponds to moving along this flat direction is the dilaton. We conclude that any effective theory that aims at describing spontaneous scale symmetry breaking (among others), must allow for a non-compact flat direction in its potential.

For instance, if we added a generic term preserving scale invariance but breaking the exchange symmetry between $|\psi|$ and $\xi$, namely (without loss of generality)
\begin{equation}\label{ex_po}
V= \lambda (|\psi|^2-\xi^2)^2 + \lambda' (|\psi|^2)^2\ ,
\end{equation}
the equations extremizing the potential would become
\begin{align}
\lambda\psi (|\psi|^2-\xi^2) &= -\lambda' |\psi|^2 \psi \ ,\\
\lambda \xi (|\psi|^2-\xi^2) &=0 \ .
\end{align}
Considering $\lambda'>0$ for $V$ to be bounded from below, the only solution is $\xi=0=\psi$, \emph{i.e.} the flat direction is completely lifted, even though scale invariance is respected.

\section{Spontaneous symmetry breaking at finite density}
\label{muuu}

In this section we depart from the Lorentz-invariant set-up discussed above, by introducing a non-zero chemical potential $\mu$ for the charge associated to the global $U(1)$ symmetry. As we will see, we will still be able to identify the dilaton and the $U(1)$ NG, though their dispersion relations will be modified in an interesting way.

We start with a scale-invariant theory defined by the action
\be
S=\int d^4x\ \left[\partial_\mu\psi^*\partial^\mu \psi+\frac{1}{2}\partial_\mu \xi\partial^\mu\xi-\lambda(|\psi|^2-\xi^2)^2-\lambda'(|\psi|^2)^2 \right] \ ,
\label{actionTOYTOYChemical}
\ee
whose potential corresponds to the extension already introduced in \eqref{ex_po}. We are going to switch on a chemical potential $\mu$ for the $U(1)$ symmetry. As discussed before, at finite chemical potential, the ground state is no longer determined by the Hamiltonian $H$ but by the effective Hamiltonian $\widetilde{H}=H-\mu Q$, where $Q$ is the $U(1)$ charge operator. As we will discuss, this modifies the effective potential of the theory and allows the fields to acquire a non-zero value. Notably, one can recover the zero chemical potential symmetry breaking case described by \eqref{tuned} by means of an appropriate limit for both $\mu$ and $\lambda'$.
The main result of the present section is to show that the dilatonic mode acquires a gap, which depends on $\mu$ and $\lambda'$.

A nonzero chemical potential can be implemented by extracting a time-dependent phase from the complex field
\be
\psi=e^{i\mu t}\phi\ ,\qquad \psi^*=e^{-i\mu t}\phi^* \ .
\label{backgroundChemical}
\ee
The equations of motion then read
\be\label{eq:eomsmu}
\partial^2 \phi+2i\mu\, \partial_0 \phi-\mu^2 \phi+\partial_{\phi^*} V(|\phi|,\xi)=0\ , \qquad
\partial^2\xi+\partial_\xi V(|\phi|,\xi)=0 \ ,
\ee
where $V(|\phi|,\xi)\equiv V(|\psi|,\xi)$ is given by \eqref{ex_po}.
Note that these equations can equivalently be obtained introducing \eqref{backgroundChemical} in \eqref{actionTOYTOYChemical}, identifying a new effective potential $V_\phi(|\phi|,\xi)=V(|\phi|,\xi)-\mu^2|\phi|^2$ and taking the variation with respect to $\phi^*$, $\xi$. Although $V_\phi$ is not the true potential (indeed, the energy density is  $E \sim V(|\phi|,\xi)+\mu^2 |\phi|^2$), the extrema of $V_\phi$ correspond to solutions of the equations of motion of the original action \eqref{actionTOYTOYChemical}. We will show in the following that $V_\phi$ determines the ground state for the effective Hamiltonian $\widetilde{H}$.

\subsection{Effective Hamiltonian and ground state}

In order to determine the effective Hamiltonian and the associated ground state we need to find expressions for the $U(1)$ charge $Q$ and Hamiltonian. We will use the usual definitions in terms of the temporal components of the energy-momentum tensor $T_{\mu\nu}$ and $U(1)$ current $J_\mu$
\be\label{eq:HQ}
H=\int d^3 x\, T_{00}\ ,\qquad Q=\int d^3 x\, J_0\ .
\ee
Then, the effective Hamiltonian at finite chemical potential is determined by the temporal component of an effective energy-momentum tensor $t_{\mu\nu}$ 
\be\label{eq:effH}
\widetilde{H}= \int d^3 x \, (T_{00}-\mu J_0)\equiv  \int d^3 x \, t_{00}.
\ee

The $U(1)$ current can be written as follows
\be
J_0=2\mu |\phi|^2+j_0\ ,\qquad J_i=j_i\ ,
\ee
where
\be\label{eq:jm}
j_\mu=i(\partial_\mu\phi^*\phi-\phi^*\partial_\mu\phi) \ .
\ee
Similarly, for the energy-momentum tensor%
\footnote{The notations are such that the capital letters ($T_{00}$ \emph{etc.}) 
refer to the dynamics of $(\psi, \xi)$ (and by extension, of $\phi$) dictated by \eqref{actionTOYTOYChemical}. 
The low case letters refer instead to the dynamics given by \eqref{eq:lagranphi} which is not the Lagrangian for $(\phi,\,\xi)$ but it shares the same potential.}
\begin{align}
&T_{00}=\mu J_0 +t_{00}, \label{TOYTOYEnergyChemical}\\
&T_{0i}=T_{i0}= \mu J_i+t_{0i}=\mu j_i+t_{0i}, \label{boo_pre}\\ 
&T_{ij}= t_{ij}+\delta_{ij}\left(\mu J_0-2\mu^2 |\phi|^2\right)=t_{ij}+\delta_{ij}\mu j_0,
\end{align}
where
\be\label{eq:tmn}
t_{\mu\nu}=2\partial_{(\mu}\phi^*\partial_{\nu)}\phi+\partial_\mu \xi\partial_\nu\xi-\eta_{\mu\nu}{\cal L}_\phi+\frac{1}{3}(\eta_{\mu\nu}\partial^2-\partial_\mu\partial_\nu)\left(\frac{1}{2}\xi^2+|\phi|^2\right),
\ee
and
\be\label{eq:lagranphi}
{\cal L}_\phi=\partial_\mu\phi^*\partial^\mu \phi+\frac{1}{2}\partial_\mu \xi\partial^\mu\xi-\lambda(|\phi|^2-\xi^2)^2-\lambda'(|\phi|^2)^2+\mu^2|\phi|^2.
\ee
Notice that from \eqref{boo_pre} we have that $T_{0i}=T_{i0}$ implying that the Ward-Takahashi identities for boost transformations 
are satisfied, so that the full Lorentz symmetry is still preserved in the presence of a non-vanishing chemical potential.

The effective potential for ${\cal L}_\phi$ is the one we had identified previously in the equations of motion \eqref{eq:eomsmu}
\be
V_\phi=\lambda(|\phi|^2-\xi^2)^2+ \lambda'(|\phi|^2)^2-\mu^2|\phi|^2,
\label{TOYTOYEffectivePot}
\ee
Since $t_{00}$ determines the effective Hamiltonian \eqref{eq:effH}, we see that the ground state will correspond to the minimum of the effective potential. The effective potential has three extrema\footnote{Note that these uniform and static solutions are extrema of the effective potential \eqref{TOYTOYEffectivePot}, but not of the energy \eqref{TOYTOYEnergyChemical}.}
\be
\xi=\phi=0\ ;\qquad
\xi=0, |\phi|^2=v^2=\frac{\mu^2}{2(\lambda+\lambda')}\ ;\qquad
\xi^2=|\phi|^2=v^2=\frac{\mu^2}{2\lambda'}\ .
\label{TOYTOYextremaChemical}
\ee
Out of the three extrema \eqref{TOYTOYextremaChemical}, the first two are saddle points and only the last is a minimum, which is the true ground state of the system. Note that for the true minimum to exist, and for $V_\phi$ to be bounded from below, we need to have $\lambda'>0$. In other words, we need to lift the flat direction that we had at $\mu=0$ in order to have a minimum, and symmetry breaking, when $\mu\neq0$.

We now proceed to investigate the low-energy spectrum around this (degenerate) minimum.

\subsection{Nambu-Goldstone dynamics from Ward-Takahashi identities}
\label{WTi}

We perturb the fields around the ground state $\xi^2=|\phi|^2=v^2=\frac{\mu^2}{2\lambda'}$. We use the same parameterization as in \eqref{psiparam}, though adapted to the field $\phi$
\be
\label{vac_flu}
\phi=e^{ i\frac{\vartheta}{\sqrt{2}v}}\left( v e^{\frac{\tau}{\sqrt{3}v}}+\frac{1}{\sqrt{6}}\rho\right)\ , \qquad
\xi= v e^{\frac{\tau}{\sqrt{3}v}}-\frac{2}{\sqrt{6}}\rho\ .
\ee
As before, the kinetic terms are diagonal and canonically normalized for $\vartheta$, $\tau$ and $\rho$. 
We still identify $\vartheta$ as the fluctuation of the phase of the condensate and $\tau$ as a 
fluctuation of its magnitude, while $\rho$ corresponds to an orthogonal direction of increasing potential energy. 
For $\mu=0$, $\vartheta$ and $\tau$ are naturally associated to the $U(1)$ NG and dilaton, while $\rho$ 
enters as a Higgs fluctuation. This simple picture is a bit complicated when $\mu\neq0$, as the would-be Goldstones undergo some mixing and also a non-vanishing gap for one linear combination. We will study this effect in some 
approximation here and in more detail in the next section.

When the perturbation \eqref{vac_flu} is introduced in the effective potential \eqref{TOYTOYEffectivePot} and expanded to quadratic order, one finds no term for $\vartheta$ and the following mass matrix for $(\tau,\rho)$
\be
M=\frac{4v^2}{3}\left(\begin{array}{cc} 2\lambda' & \sqrt{2}\lambda' \\ \sqrt{2}\lambda' & \lambda'+9{\lambda}   \end{array} \right).
\ee
In principle both perturbations are massive and mixed, but in the limit $\lambda'\ll \lambda$ in which there is an almost 
flat direction in the original potential \eqref{ex_po}, the mixing becomes very 
small and there is a large hierarchy between the mass of $\tau$, $m_\tau^2 \sim \lambda'v^2\sim\mu^2$, and the mass 
of $\rho$, $m_\rho^2\sim \lambda v^2$. In the following we will assume that we are 
in this situation, in which case the Higgs fluctuation $\rho$ can be set to zero in the low energy description to a good approximation.

The dynamical equations for the remaining fluctuations can be derived from the Ward-Takahashi identities. 
When evaluated on-shell the $U(1)$ current should be conserved and the trace of the energy momentum tensor should vanish
\be
\partial_\mu J^\mu =0\ ,\qquad
T^\mu_{\ \mu}=0.
\ee
This gives two equations, which is sufficient to determine the dynamics of $\vartheta$ and $\tau$. 
The trace of the energy-momentum tensor, to linear order in the fluctuations, is
\be
T^\mu_{\ \mu }\simeq \sqrt{3}v\left( \partial^2\tau+\frac{4}{3}\mu^2\tau-2\sqrt{\frac{2}{3}}\mu\partial_0\vartheta\right)\ ,
\ee
whereas the divergence of the current is
\be
\partial^\mu J_\mu \simeq \sqrt{2}v\left(\partial^2\vartheta +2\sqrt{\frac{2}{3}}\mu\partial_0\tau\right)\ .
\ee
This translates into the set of coupled equations
\be \label{eq:eomgold}
\begin{split}
& \partial^2\tau+\frac{4}{3}\mu^2\tau-2\sqrt{\frac{2}{3}}\mu\partial_0\vartheta\simeq 0\ ,\\
&\partial^2\vartheta +2\sqrt{\frac{2}{3}}\mu\partial_0\tau\simeq 0\ .
\end{split}
\ee
As suggested by the general analysis in the introduction, the chemical potential introduces a mixing 
between the $U(1)$ NG and the dilaton. The equations can be diagonalized using expansions in Fourier modes
\be
\tau(x^0,\mathbf{x})=\int \frac{d\omega d^3 q}{(2\pi)^4} e^{-i\omega x^0+i \mathbf{q}\cdot \mathbf{x}}\, \widetilde{\tau}(\omega,\mathbf{q})\ , \ \ \vartheta(x^0,\mathbf{x})=\int \frac{d\omega d^3 q}{(2\pi)^4} e^{-i\omega x^0+i \mathbf{q}\cdot \mathbf{x}}\, \widetilde{\vartheta}(\omega,\mathbf{q})\ . \ \
\ee
Expanding at low momentum $q^2/\mu^2 \ll 1$, the equations have solutions when the modes satisfy the dispersion relations 
\be
\omega^2 \simeq \frac{q^2}{3}\ , \qquad \omega^2\simeq 4\mu^2+\frac{5}{3}q^2.
\label{TOYTOYdisprelWIChemical}
\ee
Therefore, there is a gapless mode $\pi$ and a gapped mode $\sigma$, which at low momentum correspond respectively to the combinations
\be
\widetilde{\pi} \simeq \widetilde{\vartheta}-i\sign(\omega/q)\frac{q}{\sqrt{2}\mu}\widetilde{\tau}\ ,\qquad
\widetilde{\sigma}\simeq \widetilde{\tau}- i\sqrt{\frac{2}{3}}\sign(\omega/\mu) \left( 1+\frac{q^2}{24\mu^2}\right)\widetilde{\vartheta}.
\label{cominations}
\ee
A few comments are in order. In the first place, the dispersion relation of $\pi$ in \eqref{TOYTOYdisprelWIChemical} is such that it moves at the speed of sound as fixed by conformal invariance $c_s^2=1/3$, i.e. it can be identified as a conformal superfluid phonon, while $\sigma$ is the gapped dilaton. This identification is consistent with an effective field theory approach, see e.g.~\cite{Orlando:2019skh}.
Note that the mixing is necessary for this to happen, otherwise the phonon would move at the speed of light 
due to relativistic invariance of the rest of the terms. The second observation is that the gap of $\sigma$ is fixed by 
the chemical potential $m_\sigma=2\mu$, and independent of the couplings $\lambda$ and $\lambda'$ in this approximation. This is very reminiscent of the massive Goldstone bosons appearing when internal symmetries are spontaneously broken in the presence of a chemical potential. A last observation is that because of the mixing, it is no longer true that each Ward-Takahashi identity is tied to one specific mode. Indeed reexpressing $\tau$ and $\vartheta$ in terms of $\pi$ and $\sigma$, one can easily see that both fields appear in both equations \eqref{eq:eomgold}.

\subsection{Exact dispersion relations}

The results obtained from the Ward-Takahashi identities are easy to interpret 
physically but we had to introduce several approximations to derive them, in 
particular we used the hierarchy between the masses of the Higgs fluctuation and 
the dilaton to freeze out the first. In order to go beyond this approximation 
we need to include the Higgs mode in the analysis, whose dynamics is not 
captured by the Ward-Takahashi identities. This can be more simply done using 
the effective Lagrangian.

Consider again the vacuum $\xi^2=|\phi|^2=v^2=\frac{\mu^2}{2\lambda'}$ and the fluctuations \eqref{vac_flu} around it.
The quadratic Lagrangian for the fluctuations is 
\begin{equation}
\begin{split}
\mathcal{L}_{\text{quad}} \,=\, & \frac{1}{2}\partial_\mu\rho \partial^\mu\rho +\frac{1}{2}\partial_\mu\vartheta \partial^\mu\vartheta + \frac{1}{2}\partial_\mu\tau \partial^\mu\tau \\
& + 2 \sqrt{\frac{2}{3}}\, \mu\, \tau   \partial_t \theta + \frac{2}{\sqrt{3}} \, \mu\,\rho  \partial_t \theta 
-\frac{2}{3} \sqrt{2} \mu ^2 \tau \rho -\frac{2}{3} \mu ^2 \tau^2-\mu ^2\frac{ 9 \lambda +\lambda' }{3 \lambda'} \rho^2\ .
\end{split}
\end{equation}
By going to Fourier space we get
\begin{equation}\label{quad}
 \mathcal{L}_{\text{quad}} = \frac12\ y^T(-\omega, -q) \cdot M(\omega,q) \cdot y(\omega,q)\ , \qquad 
 y = (\vartheta, \rho, \tau)\ ,
\end{equation}
where
\begin{equation}\label{mat}
M= \left(
\begin{array}{ccc}
 \omega ^2-q^2 &i \frac{2 }{\sqrt{3}}\mu  \omega  & i\frac {2 \sqrt{2}  }{\sqrt3}\mu  \omega \\
 -i\frac{2 }{\sqrt{3}}\mu  \omega  &  \omega ^2-q^2-\frac{2(9 \lambda +\lambda')}{3
   \lambda'} \mu ^2 & -\frac{2\sqrt{2}}{3}  \mu ^2 \\
 -i \frac {2 \sqrt{2}  }{\sqrt3}\mu  \omega  & -\frac{2\sqrt{2}}{3}  \mu ^2 &  \omega ^2-q^2-\frac{4 }{3}\mu ^2 \\
\end{array}
\right)\ .
\end{equation}
Studying the zeros of the determinant of $M$, one finds one massless mode, the $U(1)$ NG boson, and two gapped modes:
\begin{equation}
\begin{split}
& \omega^2_1|_{q=0} = 0 \ , \\
& \omega^2_{2,3}|_{q=0}  = \frac{3\mu^2}{\lambda'} \left(\lambda + \lambda' \pm  \sqrt{ \lambda ^2 -\frac23 \lambda  \lambda' + \lambda'^2 }\right) \ .
\end{split}\label{3disp}
\end{equation}
Expanding for low momentum $q$ and for $\lambda'\ll \lambda$, we get
\begin{align}
 \omega_1^2 &\simeq \frac{1}{3} q^2 \ ,\\
 \omega^2_{2} &\simeq 6 \mu ^2 \frac{ \lambda }{\lambda'}\left(1+\frac{ \lambda'}{3\lambda }\right)+ \left(1+\frac{2
   \lambda'}{9 \lambda }\right)q^2 \ ,\\
 \omega^2_{3} &\simeq 4\mu^2 \left(1
 -\frac{ \lambda'}{3\lambda }\right)
 +\left(\frac{5}{3} - \frac{2 \lambda'}{9 \lambda }\right) q^2\ .
\end{align}
Comparing with the dispersion relations in \eqref{TOYTOYdisprelWIChemical}, we observe that the speed of the phonon is not modified by corrections depending on $\lambda'$, while the mass of the gapped dilaton is corrected, though mildly. Indeed,  
contrary to massive NG bosons associated to internal symmetries, the mass of the gapped dilaton is not protected by the symmetry. 

For $\lambda'\ll \lambda$, $\omega_1^2$ and $\omega^2_{3}$ reduce to the dispersion relations obtained in \eqref{TOYTOYdisprelWIChemical}
from the study of the Ward-Takahashi identities, and we have a hierarchy between the two massive modes. Furthermore, in the limit
 \begin{equation}
  \mu \rightarrow 0\ , \qquad
  \lambda' \rightarrow 0 \qquad 
  \text{with}\qquad 
  \frac{\mu^2}{2 \lambda'} \rightarrow v^2 \ ,
 \end{equation}
we recover the masses \eqref{qua_lor} of the relativistic model \eqref{tuned}
\begin{equation}
\begin{split}
& \omega^2_1|_{q=0} = 0 \ , \\
& \omega^2_{2}|_{q=0}  = 12 v^2 \lambda \ , \\
& \omega^2_{3}|_{q=0}  = 0   \ ,
\end{split}
\end{equation}
and $\omega_3$ describes the massless dilaton. This suggests a connection between 
the corrections to the mass of the gapped dilaton at finite chemical potential 
and the lack of a flat direction in the potential at zero chemical potential. 
The masses of gapped NGs might be protected only if there are flat directions 
associated to them, of course this will always be the case for internal symmetries.

\section{Boosted superfluid}
\label{vvv}

Since the chemical potential breaks Lorentz invariance, 
it is interesting to study the effect on the NG modes when the superfluid is set on motion relative 
to the frame determined by the effective Hamiltonian induced by the chemical potential, that one can identify as the ``laboratory'' frame.
We consider again \eqref{tuned} and introduce both a chemical potential and a superfluid velocity
\be
\psi=e^{i\mu_0 u_\mu x^\mu}\phi\ ,\qquad \psi^*=e^{-i\mu_0 u_\mu x^\mu}\phi^*.
\label{backgroundSuperfluid}
\ee
Where $u_\mu=\gamma(1,-\vec{\beta})$, $\gamma=1/\sqrt{1-|\beta|^2}$ is a time-like four-velocity $u_\mu u^\mu=+1$. The chemical potential is $\mu=\gamma \mu_0$, and the time direction in the laboratory frame is $x^0$. The background plane wave \eqref{backgroundSuperfluid} is the same as \eqref{backgroundChemical} seen by a boosted observer, compared to the laboratory frame.  Since \eqref{tuned} is Lorentz invariant, the dispersion relations for the gapless low-energy modes can be obtained by boosting 
those obtained from \eqref{actionTOYTOYChemical} (\emph{i.e.} the case with just a chemical potential). 
For the sake of providing an explicit check, we repeat the exercise of computing them directly through the Ward-Takahashi identities and through the perturbative Lagrangian approach. 

\subsection{Effective Hamiltonian and ground state}

We proceed in a similar fashion to the case of zero velocity. The Hamiltonian and the 
charge are still determined by the energy-momentum tensor and the current as in \eqref{eq:HQ}, 
and the effective Hamiltonian at nonzero chemical potential by \eqref{eq:effH}. Because of the boost, the expressions for the current and the energy-momentum tensor are slightly modified. 
\be
\begin{split}
&J_\mu=2\mu_0|\phi|^2u_\mu+j_\mu\ ,\\
&T_{\mu\nu}=2\mu_0^2 u_\mu u_\nu|\phi|^2+\mu_0(u_\mu j_\nu+u_\nu j_\mu)-\eta_{\mu\nu} \mu_0 u^\alpha j_\alpha+ t_{\mu\nu}(\mu_0)\ .
\end{split}
\ee
Where $j^\mu$ and $t_{\mu\nu}$ take the same form as before \eqref{eq:jm} and \eqref{eq:tmn}, replacing $\mu$ by $\mu_0$. Recalling that the chemical potential is $\mu=\mu_0 u_0=\mu_0\gamma$, the effective Hamiltonian is
\be
H-\mu Q=\int d^3x\,\left( T_{00}-\mu J_0\right)=\int d^3x\, (t_{00}(\mu_0)-\mu \vec{\beta}\cdot \vec{j}).
\ee
Since $j_i$ vanishes for constant $\phi$, the extrema of the effective potential are the same as before \eqref{TOYTOYextremaChemical} replacing $\mu$ by the effective chemical potential in the rest frame of the fluid $\mu_0$. The ground state is thus $\xi^2=|\phi|^2=v_0^2=\frac{\mu_0^2}{2\lambda'}$.

\subsection{Nambu-Goldstone dynamics from Ward-Takahashi identities}

We can use the same parametrization for perturbations of the ground state as in \eqref{vac_flu}, replacing $v$ by $v_0$. The same considerations about the mass hierarchy of $\tau$ and $\rho$ apply, so in this analysis we will assume $\lambda'\ll \lambda$ and freeze $\rho$. The dynamics of the low energy modes are determined by the conservation equations for the current and the energy-momentum tensor. For the boosted superfluid they take the form
\be
\begin{split}
&\partial^\mu J_\mu=2\mu \left(\partial_0+\vec{\beta}\cdot\vec{\nabla}\right)|\phi|^2+\partial^\mu j_\mu,\\
&T^\mu_{\ \mu}=2\mu_0^2|\phi|^2-2\mu_0 u^\mu j_\mu+t^\mu_{\ \mu}(\mu_0)=2\mu_0^2|\phi|^2-2\mu(j_0+\vec{\beta}\cdot \vec{j})+t^\mu_{\ \mu}(\mu_0).
\end{split}
\ee
Therefore, we should just replace the terms with a single time derivative by the material derivative 
$\mu\partial_0\to \mu D_0=\mu (\partial_0+\vec{\beta}\cdot \vec{\nabla})$ and otherwise change $\mu$ by the effective $\mu_0$:
\be
\begin{split}
&\partial^\mu J_\mu \simeq \sqrt{2} v_0\left(\partial^2\vartheta +2\sqrt{\frac{2}{3}}\mu D_0\tau \right),\\
&T^\mu_{\ \mu }\simeq \sqrt{3}v_0\left( \partial^2\tau+\frac{4}{3}\mu_0^2\tau-2\sqrt{\frac{2}{3}}\mu D_0\vartheta\right).
\end{split}
\ee
From this, we obtain the equations
\be\label{eq:labframeeqs}
\begin{split}
& \partial^2\tau+\frac{4}{3}\mu_0^2\tau-2\sqrt{\frac{2}{3}}\mu D_0\vartheta\simeq 0\ ,\\
&\partial^2\vartheta +2\sqrt{\frac{2}{3}}\mu D_0\tau\simeq 0\ .
\end{split}
\ee

The dispersion relation for the gapless mode can be more easily found by noting that $\mu=\gamma\mu_0$ and using comoving coordinates. Taking $\vec{\beta}$ parallel to the $x^3$ direction, we introduce 
\be
x^0=\gamma (x^0_\beta+\beta x^3_\beta)\ ,\qquad
x^3=\gamma( x_\beta^3+\beta x^0_\beta)\ ,\qquad
x^1=x^1_\beta\ ,\qquad
x^2=x^2_\beta.
\ee
Then
\be
\frac{\partial}{\partial x^0_\beta}=\gamma (\partial_0+\beta \partial_3)\ ,\qquad 
\frac{\partial}{\partial x^3_\beta}=\gamma (\partial_3+\beta \partial_0)\ ,\qquad \partial^2=\partial^2_\beta\ .
\ee
The equations become
\be
\begin{split}
& \partial_\beta^2\tau+\frac{4}{3}\mu_0^2\tau-2\sqrt{\frac{2}{3}}\mu_0 \partial_{x^0_\beta}\vartheta\simeq 0,\\
&\partial_\beta^2\vartheta +2\sqrt{\frac{2}{3}}\mu_0 \partial_{x^0_\beta}\tau\simeq 0.
\end{split}
\ee
These are the same as before \eqref{eq:eomgold}, replacing $\mu$ by $\mu_0$. We introduce an expansion of the modes in the rest frame in plane waves  
\be
\begin{split}
&\tau(x^0_\beta,\mathbf{x}_\beta)=\int \frac{d\omega_\beta d^3 q_\beta}{(2\pi)^4} e^{-i\omega_\beta x^0_\beta+i \mathbf{q}_\beta\cdot \mathbf{x}_\beta}\, \widetilde{\tau}(\omega_\beta,\mathbf{q}_\beta), \\ &\vartheta(x^0_\beta,\mathbf{x}_\beta)=\int \frac{d\omega_\beta d^3 q_\beta}{(2\pi)^4} e^{-i\omega_\beta x^0_\beta+i \mathbf{q}_\beta\cdot \mathbf{x}_\beta}\, \widetilde{\vartheta}(\omega_\beta,\mathbf{q}_\beta). \ \
\end{split}
\ee
We recover the expected low momentum dispersion relations in the rest frame
\be\label{eq:disprestframe}
\omega_\beta^2\simeq c_s^2 q_\beta^2\ , \qquad 
\omega_\beta^2 \simeq  4\mu_0^2+\frac{5}{3} q_\beta^2\ ,
\ee 
where $c_s^2=1/3$ is the speed of sound of the scale invariant theory. 
These expressions can be translated to frequency and momentum in the laboratory frame using that
\be
\omega=\gamma(\omega_\beta+\beta q_{\beta\, 3})\ ,\qquad q_3=\gamma (q_{\beta\, 3}+\beta \omega_\beta)\ ,\qquad q_1=q_{\beta\, 1}\ ,\qquad q_2=q_{\beta\, 1}.
\ee
Note that the dispersion relations \eqref{eq:disprestframe} are valid for low momentum in 
{\em the rest frame of the fluid} $|q_\beta| \ll |\mu_0|$. For the gapless modes they can be matched with a low momentum expansion in the laboratory frame $|q|\ll |\mu|$, however for the gapped modes this is not possible, as for generic $\beta$, $q_3\sim   \omega_\beta \sim \mu$. Therefore, finding the dispersion relations of the gapped modes at low momentum {\em in the laboratory frame} requires solving \eqref{eq:labframeeqs} directly.

We classify the dispersion relations of the gapless modes taking as reference the direction of the superfluid velocity in the laboratory frame. The dispersion relations for the longitudinal modes is
\be
\label{c_lon}
\omega_\parallel=\pm\frac{ c_s\pm \beta}{1\pm \beta c_s} q_3\ , \ \ q_1=q_2=0\ ,
\ee
while the dispersion relation for the transverse modes is 
\be
\omega_\perp^2= c_s^2 \frac{q_1^2+q_2^2}{\gamma^2(1-\beta^2 c_s^2)}\ ,\qquad q_3=0\  \label{c_tra_massless}.
\ee
These expressions agree with the ones obtained by relativistic addition of velocities. Note that for $|\beta| > c_s$ both (positive frequency) longitudinal modes \eqref{c_lon} propagate in the same direction as the superfluid velocity. This is the reason why we expressed linearly the dispersion relations.

For the gapped modes the low momentum dispersion relations are
\be\label{eq:gapmodebeta}
\begin{split}
& \omega_\parallel =  2 \sqrt{1-\beta^2 c_s^2}\mu
-\frac{2}{3}\frac{\beta q_3}{1-\beta^2 c_s^2} +\frac{5+\beta^2c_s^2}{12\gamma^2(1-\beta^2 c_s^2)^{5/2}}\frac{q_3^2}{\mu}\ 
,\ \ q_1=q_2=0,\\
&\omega_\perp^2= 4 (1-\beta^2 c_s^2)^2\mu^2 +\frac{5-\beta^2}{3(1-\beta^2 c_s^2)}(q_1^2+q_2^2)\ ,\ \ q_3=0\ ,
\end{split}
\ee
where again the longitudinal dispersion relation is expressed linearly.
The gap is reduced by the superfluid velocity, but in this approximation remains finite even in the limit $\beta\to 1$, where the condensate vanishes (i.e.~at fixed $\mu$). 
Note also that at leading order for momenta in the same direction of the flow, the frequency is reduced.

\subsection{Exact dispersion relations}

We now study the effects of including the Higgs fluctuation $\rho$, and the corrections for finite $ \lambda'/\lambda$. We thus resort to expanding the full Lagrangian.
According to \eqref{backgroundSuperfluid}, we switch on a chemical potential $\mu= \mu_0\gamma$ and a background wave vector $k_3 = \mu_0 \gamma \beta$. The effective potential is now:
\begin{equation}
V= \lambda (|\phi|^2-\xi^2)^2 + \lambda' |\phi|^4-(\mu^2 - k_3^2)|\phi|^2\ ,
\end{equation}
For stationary solutions, the situation is not very different from the case with just $\mu$,
in fact one just needs to replace $\mu^2$ by $(\mu^2 - k_3^2)=\mu_0^2$ in \eqref{TOYTOYEffectivePot}. Therefore, we consider the solution
\begin{equation}\label{solv}
\xi^2=|\phi|^2\ , \qquad \qquad |\phi|^2 = \frac{\mu^2_0}{2 \lambda'}\ .
\end{equation}
The fluctuation around \eqref{solv} are still given by \eqref{vac_flu}
where however $v=v_0=\frac{\mu_0}{\sqrt{2 \lambda'}}$.
Writing $k_3=\beta\mu$, the quadratic Lagrangian for the fluctuations is
\begin{equation}
\begin{split}
\mathcal{L}_{\text{quad}} \,=\, & \frac{1}{2}\partial_\mu\rho \partial^\mu\rho 
+\frac{1}{2}\partial_\mu\vartheta \partial^\mu\vartheta 
+ \frac{1}{2}\partial_\mu\tau \partial^\mu\tau \\
& + 2 \sqrt{\frac{2}{3}}\, \mu\, \tau   (\partial_t+\beta \partial_3 )\theta 
+ \frac{2}{\sqrt{3}} \, \mu\,\rho  (\partial_t+\beta \partial_3 ) \theta 
 \\
& -\frac{2}{3} \sqrt{2}\, \mu^2_0 \rho\,  \tau
-\frac{2}{3} \mu^2_0 \tau^2
-\mu^2_0 \frac{9 \lambda +\lambda' }{3 \lambda'} \rho^2\ .
\end{split}
\end{equation}
In analogy to \eqref{quad} and \eqref{mat}, by going to Fourier space we get the kinetic matrix:
\begin{equation}\label{maga}
\left(
\begin{array}{ccc}
\omega ^2-q^2 &i \frac{2 }{\sqrt{3}}\mu  \left(\omega-\beta q_3\right)  & i\frac {2 \sqrt{2}  }{\sqrt3}\mu  \left(\omega-\beta q_3\right) \\
-i\frac{2 }{\sqrt{3}}\mu  \left(\omega-\beta q_3\right)  &  \omega ^2-q^2-\frac{2(9 \lambda +\lambda')}{3
	\lambda'} \mu^2_0 & -\frac{2\sqrt{2}}{3}  \mu^2_0 \\
-i \frac {2 \sqrt{2}  }{\sqrt3}\mu  \left(\omega-\beta q_3\right)  & -\frac{2\sqrt{2}}{3}  \mu^2_0 &  \omega ^2-q^2-\frac{4 }{3}\mu^2_0 \\
\end{array}
\right)\ .
\end{equation}
From the determinant of \eqref{maga}, one can find the exact dispersion relations. First of all, setting the momenta $q=0$ one finds that there is a massless mode corresponding to the $U(1)$ NG boson and two gapped modes:
\begin{equation}
\begin{split}
& \omega^2_1|_{q=0} = 0 \ , \\
& \omega^2_{2,3}|_{q=0}  = \frac{3}{ \lambda'}\bigg[\lambda \mu^2_0 +\lambda' \mu^2(1-c_s^2\beta^2) 
\pm\sqrt{\lambda^2 \mu^4_0-\frac23\lambda \mu^2_0\lambda' \mu^2(1-c_s^2\beta^2)+\lambda'^2 \mu^4(1-c_s^2\beta^2)^2}\bigg] ,
\end{split}\label{gapsfull}
\end{equation}
where $c_s=1/3$ as before.

Now, expanding at low frequencies and momenta, one can extract analytically the dispersion relation for the $U(1)$ NG mode:
\begin{equation}
\omega_1 =\frac{c_s}{1-c_s^2\beta^2}\left(2c_s \beta q_3  \pm\sqrt{(1-\beta^2)^2q_3^2+(1-\beta^2)(1-c_s^2\beta^2)(q_1^2+q_2^2)}\right) \ .
\end{equation}
Notice that the above expression is independent of the ratio $\lambda'/\lambda$. Indeed one can check that in the longitudinal and transverse case, it reproduces correctly the expressions \eqref{c_lon} and \eqref{c_tra_massless}, respectively.

For the massive modes, one has to expand the frequencies around the respective gaps. To first order in momenta and in $\lambda'/\lambda$, the dispersion relations for the  gapped dilaton are:
\begin{equation}
\begin{split}
&\omega^{(\text{gapped})}_\parallel =
 2 \mu \sqrt{1-c_s^2\beta^2} \left[1 - \frac{\lambda'}{6\lambda}\, \frac{1-c_s^2\beta^2}{1-\beta^2} \right]  - \frac{2 \beta}{3(1-c_s^2\beta^2)}  \left[1 - \frac{\lambda'}{3\lambda}\, \frac{1-c_s^2\beta^2}{1-\beta^2}\right] \ q_3+\dots\\
&\omega^{(\text{gapped})}_\perp =
2 \mu \sqrt{1-c_s^2\beta^2} \left[1 - \frac{\lambda'}{6\lambda}\, \frac{1-c_s^2\beta^2}{1-\beta^2} \right]  + \frac{1}{12\sqrt{1-c_s^2\beta^2}} \left[ \frac{5-\beta^2}{1-c_s^2\beta^2} +  \frac{\lambda'}{6\lambda} \right] \frac{q_1^2+q_2^2}{\mu}+\dots
\end{split}
\end{equation}
For $\lambda'/\lambda=0$ they agree with the dispersion relations obtained from the Ward-Takahashi identities \eqref{eq:gapmodebeta}. Again, we observe that the gap receives corrections in $\lambda'/\lambda$, so it is not protected by the symmetry.

Finally, we present the dispersion relations of the Higgs mode, to the same order:
\begin{equation}
 \begin{split}
 &\omega_\parallel^{\text(\text{heavy})} =  \sqrt{\frac{6\lambda}{\lambda'}} \sqrt{1-\beta^2}\ \mu
 \left[1 + \frac{\lambda'}{6\lambda} \frac{1-c_s^2\beta^2}{1-\beta^2}  \right]-\frac{2\lambda'}{9\lambda}\frac{\beta  }{1-\beta^2} \ q_3+\dots\\
 &\omega_\perp^{\text(\text{heavy})} =  \sqrt{\frac{6\lambda}{\lambda'}} \sqrt{1-\beta^2}\ \mu
 \left[1 + \frac{\lambda'}{6\lambda} \frac{1-c_s^2\beta^2}{1-\beta^2}  \right]+ \frac{1}{2\sqrt{1-\beta^2}}\sqrt{\frac{\lambda'}{6\lambda}} \ \frac{q_1^2+q_2^2}{\mu}+\dots
 \end{split}
\end{equation}
Note that the $\beta\to1$ limit at fixed $\mu$ seems to be ill-defined. However this is an artifact of the expansion. For instance, inspecting \eqref{gapsfull} and noticing that in this limit $\mu_0\to0$, we find that the gap of the dilaton actually goes to zero, while the gap of the Higgs mode stays finite, but scales with $\mu$, which might be slightly non-intuitive (recall that in this limit there is no condensate).

\section{Summary and discussion}
\label{disco}

The two main highlights of the present paper are:
\begin{enumerate}
 \item The analysis of the low-energy mode associated to spontaneously 
 broken scale symmetry and the characterization of how its zero-temperature 
 gap depends on the finite density.
 \item The description of a generic method based on Ward-Takahashi identities alone to 
study the low-energy modes of an effective field theory.
\end{enumerate}

The analysis pursued in the present paper indicates that the spontaneous breaking of the scale symmetry
at zero temperature gives rise to a light dilatonic mode whose gap is directly proportional to the chemical potential.
This generic expectation can be relevant for the low-energy content of zero-temperature systems 
where the chemical potential, too, is small with respect to the UV cut-off of the effective description
(related to some other physical scale such as an external magnetic field \cite{Hayes2016ScalingBM}).

Ward-Takahashi identities in quantum field theory are known to be a key tool for 
the study of symmetries, either when these are preserved or broken, 
and even when the breaking is explicit \cite{Weinberg:1996kr,Weinberg:1972fn}.
The present paper stresses that Ward-Takahashi identities alone provide a sufficient 
framework to study the dispersion relations of the low-energy modes of an 
effective field theory, providing an alternative --generally simpler-- 
approach than the direct fluctuation analysis at the level of the 
Lagrangian. The method is generic, but we applied it to the specific 
study of scale symmetry breaking to the purpose of elucidating the 
characteristics of the resulting low-energy dynamics.
It would be interesting to look for a gapped dilaton in a strongly coupled theory, by means of the holographic duality. This might be achieved by combining holographic models with a gapless dilaton in Poincar\'e invariant vacua \cite{Bajc:2013wha,Hoyos:2013gma} (see also \cite{Argurio:2014rja}) and models with type II and gapped NG modes \cite{Amado:2013xya,Argurio:2015via}.

We first examined a relativistic field-theory model \eqref{tuned} 
in four spacetime dimensions where scale symmetry and a global 
$U(1)$ symmetry are concomitantly and spontaneously broken. 
The scale-invariant potential must have two flat directions which 
translate into two gapless NG modes, the dilaton and the $U(1)$ NG both relativistic and both propagating at the 
speed of light, \eqref{EOMPhasonToyToy} and \eqref{EOMDilatonToyToy}.

In order to realize the same symmetry-breaking pattern at finite density,
the model must be stabilized by means of an extra scale-invariant 
term \eqref{actionTOYTOYChemical} which lifts the dilatonic flat 
direction without affecting the spontaneous nature of the breaking.
The resulting low-energy modes are nonetheless altered: the $U(1)$ 
NG remains gapless but propagates at the conformal speed of sound, like a superfluid phonon;
the dilaton acquires a gap of the order of the chemical potential $\mu$
whose value is however not protected by symmetry \eqref{TOYTOYdisprelWIChemical}. The dilaton is light compared to other gapped modes only when the coefficient of the term that lifts the flat direction \eqref{actionTOYTOYChemical} is tuned to be very small, in that case we observe that the dilaton gap becomes independent of the couplings.

Our results at nonzero density belong to the line of research on gapped NG modes
\cite{Kapustin:2012cr,Nicolis_2013,Watanabe:2013uya,Nicolis:2013sga,Cuomo:2020gyl}.
In this context, a natural future perspective is to embed the present analysis into a 
systematic Maurer-Cartan effective framework, thus assessing its universality and
possible generalizations.

One interesting field of applications is provided by condensed matter. The presence of a wide critical region in the phase diagram is a characteristic shared by 
many --generally strongly correlated-- systems, among which the cuprates. 
The critical phase is associated to interesting phenomena like \emph{bad} and \emph{strange metallicity} 
and non-Fermi liquid behavior \cite{Hussey_2008}. It is also often conjectured to lie at the basis 
of the mechanism for high-temperature superconductivity, see for instance \cite{Leong_2018}.

The defining property of such critical region is the validity of simple scaling rules 
whose origin, however, can involve complicated and often elusive dynamics related to the presence 
of a quantum critical point \cite{Sachdev:2011cs,Senthil:2004aza,Sachdev_2010} or, more generally,
to the presence of a scaling sector \cite{Georgi_2007,Leong_2018}. 
This is sometimes referred to as \emph{generic scale invariance} 
\cite{RevModPhys.77.579} and can be assumed among the defining symmetries of an effective description.

Another paradigmatic example is provided by cold atoms at unitarity, where there is an emergent non-relativistic conformal symmetry \cite{Nishida:2010tm}, known as Schroedinger symmetry. Gapped NG modes are known to appear when the Hamiltonian is deformed by some of the symmetry generators of the Schroedinger algebra \cite{Ohashi:2017vcy}.  An extension of our analysis, along the lines of \cite{Arav_2017}, to systems with Galilean rather than Lorentz invariance would be quite interesting.

As another remark, still related to condensed matter but in the context of standard metals,
it is relevant to mention that the low-energy modes of our analysis would not destabilize 
a Landau-Fermi liquid coexisting with them. This can be appreciated 
by means of an extension of the results of \cite{Watanabe:2014hca} to dilatations, a symmetry
which does not commute with either spatial or temporal translations: one can show that 
the linear interaction term between the fermionic quasiparticles and, respectively, the 
$U(1)$ NG and the dilaton are both vanishing.

The model adopted here allows for generalizations in which the $U(1)$ symmetry is coupled to translations 
and the symmetry-breaking preserves only a linear combination of the two \cite{Musso:2018wbv}. 
This would realize a spatial version of the pattern described above when $\mu\neq 0$ and only a diagonal component of the product 
of internal $U(1)$ and time translations was preserved. Such breakings are referred to as 
\emph{homogeneous} because they do not yield any spacetime modulation of the energy density,%
\footnote{An example of inhomogeneous breaking of spatial translations in field theory was studied in \cite{Musso:2019kii}.
Studying the cubic polynomial in $\omega^2$ associated to \eqref{mat}, one 
can exclude the presence of complex solutions. Similarly, a numerical study of \eqref{maga} showed no hints of finite-momentum instabilities. Thus, for the purpose of studying translation symmetry breaking, the models introduced in the main text need to be enriched and generalized.}
they however provide acoustic phonon modes. It is an interesting open question to study whether and how these phonons 
would coexist with a dilatonic mode \cite{wip}. The relevance of the question is three-fold: it relates to the counting
problem of NG modes for spacetime symmetries \cite{Low_2002,Watanabe_2020}; it concerns condensed matter systems where a critical 
scaling and the breaking of translations are intertwined;%
\footnote{Such as in the region of the phase diagram of cuprates overlapping with the critical, strange metal phase 
and the so-called pseudo-gap phase \cite{Hussey_2008,Delacretaz:2016ivq}.} it provides insight regarding holographic models
where scaling and translation symmetries are broken together \cite{Amoretti:2016bxs,Esposito:2017qpj,Baggioli:2019elg,Alberte:2017oqx}.%
\footnote{Merging the last two points, there is a current in the literature addressing the critical breaking 
of translations relevant for the study of strongly-correlated electron systems (specifically \emph{strange} and \emph{bad metals}),
see for instance \cite{Andrade:2013gsa,Donos:2013eha,Amoretti:2017frz,Amoretti:2017axe,Donos:2018kkm,Baggioli:2019jcm,Amoretti:2019cef,Amoretti:2019kuf}.}

\section*{Acknowledgments}

We are grateful to Tomas Brauner for useful comments. R.A. and C.H. want to thank Nordita for their hospitality during the program ``Effective Theories of Quantum Phases of Matter''. R.A.~and D.N.~acknowledge support by IISN-Belgium (convention 4.4503.15) and by the F.R.S.-FNRS under the ``Excellence of Science" EOS be.h project n.~30820817. R.A.~is a Research Director of the F.R.S.-FNRS (Belgium).
C.H.~has been partially supported by the Spanish grant PGC2018-096894-B-100 and by the Principado de Asturias through the grant GRUPIN-IDI/2018 /000174.

\bibliography{toy2} 
\bibliographystyle{utphys}

\end{document}